\documentclass[reprint,aps,prl,floatfix,superscriptaddress]{revtex4-2}

\usepackage{amsmath,amssymb,bm,caption,setspace,xspace,fancyvrb,scrextend,siunitx,euscript}
\usepackage{enumerate,braket,float,wrapfig,booktabs}
\usepackage[displaymath]{lineno}
\usepackage[pdftex]{graphicx}
\usepackage{textcomp}

\newcommand{\uu}[1]{\ensuremath{\, \mathrm{#1}}}

\usepackage{color}

\graphicspath{{./Figures/}}

\DeclareMathOperator{\Tr}{Tr}

\begin{document}

\title{Experimental limit on non-linear state-dependent terms in quantum theory}

\author{Mark Polkovnikov}
\affiliation{BU Academy, Boston, MA 02215, USA}
\author{Alexander~V.~Gramolin}
\affiliation{Department of Physics, Boston University, Boston, MA 02215, USA}
\author{David E. Kaplan}
\affiliation{Department of Physics \& Astronomy, The Johns Hopkins University, Baltimore, MD 21218, USA}
\author{Surjeet Rajendran}
\affiliation{Department of Physics \& Astronomy, The Johns Hopkins University, Baltimore, MD 21218, USA}
\author{Alexander~O.~Sushkov}
\email{asu@bu.edu}
\affiliation{Department of Physics, Boston University, Boston, MA 02215, USA}
\affiliation{Department of Electrical and Computer Engineering, Boston University, Boston, MA 02215, USA}
\affiliation{Photonics Center, Boston University, Boston, MA 02215, USA}

	

\begin{abstract}
We report the results of an experiment that searches for causal non-linear state-dependent terms in quantum field theory. Our approach correlates a binary macroscopic classical voltage with the outcome of a projective measurement of a quantum bit, prepared in a coherent superposition state. Measurement results are recorded in a bit string, which is used to control a voltage switch. Presence of a non-zero voltage reading in cases of no applied voltage is the experimental signature of a non-linear state-dependent shift of the electromagnetic field operator. We implement blinded measurement and data analysis with three control bit strings. Control of systematic effects is realized by producing one of the control bit strings with a classical random-bit generator. The other two bit strings are generated by measurements performed on a superconduting qubit in an IBM Quantum processor, and on a $^{15}$N nuclear spin in an NV center in diamond. 
Our measurements find no evidence for electromagnetic quantum state-dependent non-linearity. We set a bound on the parameter that quantifies this non-linearity $|\epsilon_{\gamma}|<4.7\times 10^{-11}$, at 90\% confidence level.
Within the Everett many-worlds interpretation of quantum theory, our measurements place limits on the electromagnetic interaction between different branches of the universe, created by preparing the qubit in a superposition state.
\end{abstract}

\maketitle

Quantum mechanics has proven to be a very successful theory of physics at microscopic scales. It describes the behavior of systems at the nuclear and atomic scale, as well as properties of materials and radiation. 
There are two fundamental principles at the foundation of quantum theory: the concept of probabilistic, rather than deterministic measurement, and the requirement of linear time evolution.
The often counter-intuitive predictions of quantum theory have generated numerous scientific debates and interpretations, as well as a number of theoretical modifications~\cite{RevModPhys.76.1267, RevModPhys.85.471, Wharton2020}. 
Nevertheless, it has withstood all experimental tests~\cite{Bell1964, PhysRevLett.63.1031, PhysRevLett.64.2261, PhysRevLett.64.2599, PhysRevLett.65.2931, Shadbolt2014, Hensen2015, Donadi2020}. Quantum field theory forms the foundation of modern physics. 

Theoretical attempts of introducing non-linear time evolution into quantum theory have generically suffered from causality problems~\cite{PhysRevLett.62.485, WEINBERG1989336, Polchinski:1990py,Gisin1990}. However, a recent theory proposal has introduced a causal mechanism for describing non-linear time evolution within the field theory framework by shifting bosonic field operators by a small amount, proportional to their expectation value in the full quantum state $|\psi\rangle$ (note that we are in the Heisenberg picture)~\cite{Kaplan2021_}. 
For example, in linear quantum electrodynamics the interaction between the electromagnetic field, given by the 4-vector potential $A_{\mu}$, and a current 4-vector $J^{\mu}$ is given by the Lagrangian $A_{\mu} J^{\mu}$.
In the proposed non-linear theory this electromagnetic interaction is modified to $\left(A_{\mu}+\epsilon_{\gamma}\langle\psi|A_{\mu}|\psi\rangle\right) J^{\mu}$,
where $\epsilon_{\gamma}$ is the parameter that quantifies the degree of non-linearity for electromagnetic fields, and we keep only lowest-order terms.
The current $J^{\mu}$ thus effectively interacts with the vector potential:
\begin{align}
A'_{\mu} = A_{\mu}+\epsilon_{\gamma}\langle\psi|A_{\mu}|\psi\rangle.
\label{eq:100}
\end{align}
This modification preserves causality, energy conservation, and gauge-invariance of the theory, as well ensuring that quantum states have a conserved norm~\cite{Kaplan2021_}.
The same non-linear construction could be extended to gravitational fields by modifying the $g_{\mu\nu}$ metric tensor as $g'_{\mu\nu} = g_{\mu\nu}+\epsilon_g\langle\psi|g_{\mu\nu}|\psi\rangle$.
This opens up a number of intriguing prospects, including the possibility to solve the black hole information problem~\cite{Marolf_2017, Kaplan2021_}. Nevertheless, current experimental constraints on such non-linear modifications are weak. The strongest bounds $\epsilon_{\gamma}\lesssim 10^{-5}$ can be deduced from ion trapping experiments, and the experimental bounds on $\epsilon_g$ are even weaker~\cite{Kaplan2021_}. 

\begin{figure}[t]
\centering
		\includegraphics[width=\columnwidth]{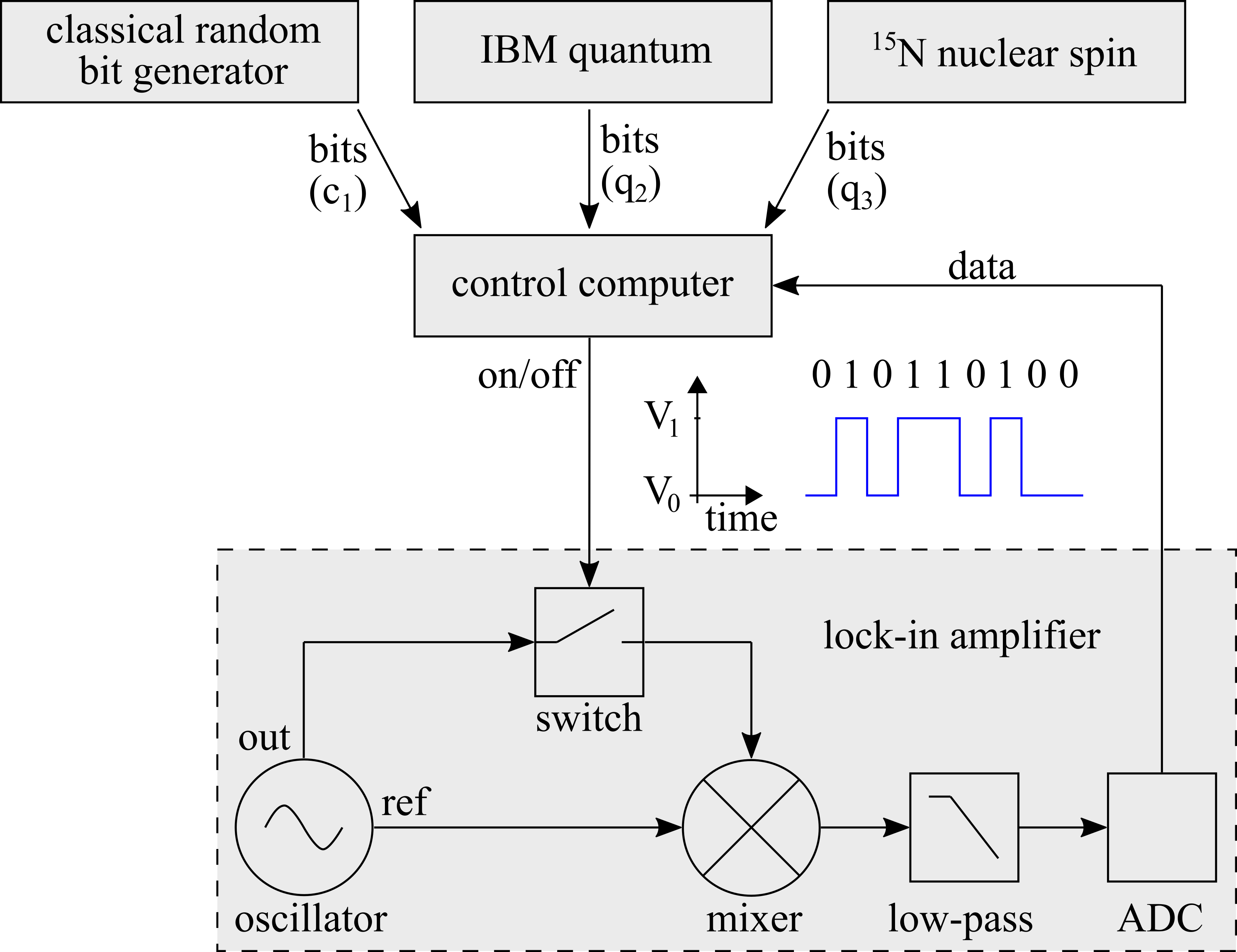}
	\caption{
		Experimental setup. Three bit strings $c_1,q_2,q_3$ were randomly permuted and used to manipulate the switch that controls the output voltage of the Zurich Instruments MFLI lock-in amplifier. 
		The lock-in output was set to $3\uu{V}$ amplitude at $1\uu{MHz}$ carrier frequency. The low-pass filter time constant was set to $1\uu{ms}$. 
		The figure shows an example of a 9-bit control string and the corresponding voltage waveform.
	}
	\label{fig:1}
\end{figure}
We report experimental limits on the electromagnetic nonlinearity parameter $\epsilon_{\gamma}$. Our approach is similar to the ``Everett phone'' proposed in ref.~\cite{Polchinski:1990py}. We correlate a macroscopic (classical) voltage with the outcome of the measurement of a quantum 2-level system (qubit). We denote the stationary states of the qubit as $|0\rangle$ and $|1\rangle$. The classical apparatus that creates and detects the voltage is the Zurich Instruments MFLI lock-in amplifier, whose oscillator output is connected directly to its voltage input, fig.~\ref{fig:1}. Our measurement procedure is as follows. (1) Initialize the qubit in state $|\chi\rangle = (|0\rangle+|1\rangle)/\sqrt{2}$. (2) Perform a qubit measurement and record the resulting qubit state. (3) Set the lock-in output voltage amplitude according to the following rule: output voltage $V_0 = 0\uu{V}$, if the qubit measurement detects it in state $|0\rangle$; output voltage $V_1 = 3\uu{V}$, if the qubit measurement detects it in the state $|1\rangle$. (4) Record the resulting voltage reading. 

Our system is described by the density matrix
$\rho = (\ket{0}\bra{0}\rho_0 + \ket{1}\bra{1}\rho_1)/2$, where $\rho_0$ and $\rho_1$ are the density matrices that describe the classical environment, and corresponding to the two possible voltage readings $\Tr{(\rho_0\hat{V})}=V_0=0$ and $\Tr{(\rho_1\hat{V})}=V_1=3\uu{V}$, where $\hat{V}$ is the voltage operator.
We emphasize that the classical apparatus is not being observed to be in a coherent superposition state~\cite{Zurek2003,Zurek2009}.
The non-linear modification, introduced in eq.~\eqref{eq:100}, modifies the voltage operator to 
\begin{align}
\hat{V}' = \hat{V} + \epsilon_{\gamma}\Tr{(\rho\hat{V})} = \hat{V} + \epsilon_{\gamma}V_1/2.
\label{eq:200}
\end{align}
Now, to lowest order in $\epsilon_{\gamma}$, the two possible voltage measurement outcomes are $\Tr{(\rho_0\hat{V})}=\epsilon_{\gamma}V_1/2$ when the observer records $\ket{0}$, and $\Tr{(\rho_1\hat{V})}=V_1=3\uu{V}$ when the observer records $\ket{1}$.
Thus we search for a non-zero voltage reading, proportional to $\epsilon_{\gamma}$, in cases when the qubit is measured to be in state $\ket{0}$.
Note that in the Everett many-worlds interpretation the trace in Eq.~\eqref{eq:200} is over the entire density matrix~\cite{Everett1957}.
In the Copenhagen interpretation the quantum state collapses after the measurement in step (2) of our procedure, and this effect vanishes. Therefore the non-linear modification of bosonic operators creates an opportunity to distinguish between the Copenhagen and the many-worlds interpretations of quantum theory, and search for the existence of other worlds created by quantum measurements.

The lock-in amplifier is controlled by a computer program, which takes as inputs three bit strings. The first bit string $c_1$ consists of $60000$ classical bits, generated by a random bit generator (implemented in Matlab R2021a software). The second bit string $q_2$ consists of $30000$ bits, generated by the cloud-based IBM Quantum processor. To obtain each bit in $q_2$, a single transmon superconducting qubit was initialized into a superposition state $|\chi\rangle = (|0\rangle+|1\rangle)/\sqrt{2}$ and measured, assigning the bit value 0 to measurement result $|0\rangle$ and bit 1 to measurement result $|1\rangle$.
The third bit string $q_3$ consists of $10717$ bits, generated by a $^{15}$N nuclear spin $I=1/2$ qubit that is part of a single NV center in a diamond crystal in our laboratory. We label the quantum states of this qubit as $|\hspace{-0.2em}\downarrow\rangle,\,|\hspace{-0.2em}\uparrow\rangle$, indicating the sign of the projection of the nuclear spin along the NV center axis.
To obtain each bit in $q_3$, the $^{15}$N nuclear spin was initialized into the $|\chi\rangle = (|\hspace{-0.2em}\downarrow\rangle+|\hspace{-0.2em}\uparrow\rangle)/\sqrt{2}$ state, and measured, assigning the bit value 0 to measurement result $|\hspace{-0.2em}\downarrow\rangle$ and bit 1 to measurement result $|\hspace{-0.2em}\uparrow\rangle$.

We use the three bit strings to control systematic errors in our experiment. The most important systematic is leakage of the lock-in oscillator output through the open switch, which leads to a non-zero open-switch detected voltage $V_s$, mimicking the signal due to quantum non-linearity. We control this systematic by checking the voltage detected when classical bits are used to control the switch. A statistically-significant difference in the detected voltage between classical and quantum bit control is a signature of quantum non-linearity $\epsilon_{\gamma}$. 
The two quantum measurement-generated bit strings $q_2$ and $q_3$ are a mechanism for incorporating the effects of classical noise and qubit readout fidelity. In this context we define readout fidelity $f$ as the probability of measuring the qubit to be in the same quantum state to which it had just been initialized~\cite{som}. Classical readout noise degrades this fidelity, and random bits generated by a classical computer correspond to $f=1/2$.
For readout fidelity $1/2\leq f\leq 1$, and in presence of an open-switch leakage voltage $V_s\ll V_1$, our experiment's possible voltage outcomes are $V_s+\epsilon_{\gamma}V_1(f-1/2)$ when the observer records $\ket{0}$, and $V_1=3\uu{V}$ when the observer records $\ket{1}$.

The fidelity of the IBM Quantum measurements that generated the $q_2$ bit string was 99\%~\cite{som}. The imperfect readout of the NV center severely degrades the fidelity of direct $^{15}$N nuclear spin measurements, which are limited by photon shot noise. To circumvent this limitation, we implemented the repetitive readout scheme, which made use of a narrow-band selective RF pulse to realize the CNOT gate between the NV center nuclear and the electron spin qubits, with subsequent projective optical measurement of the electron spin state~\cite{Jiang2009,Neumann2010,Lovchinsky2016}. Each $^{15}$N nuclear spin measurement consisted of 50 repetitive readout cycles, achieving 55\% readout fidelity~\cite{som}.

\begin{figure}[t]
	\centering
	\includegraphics[width=\columnwidth]{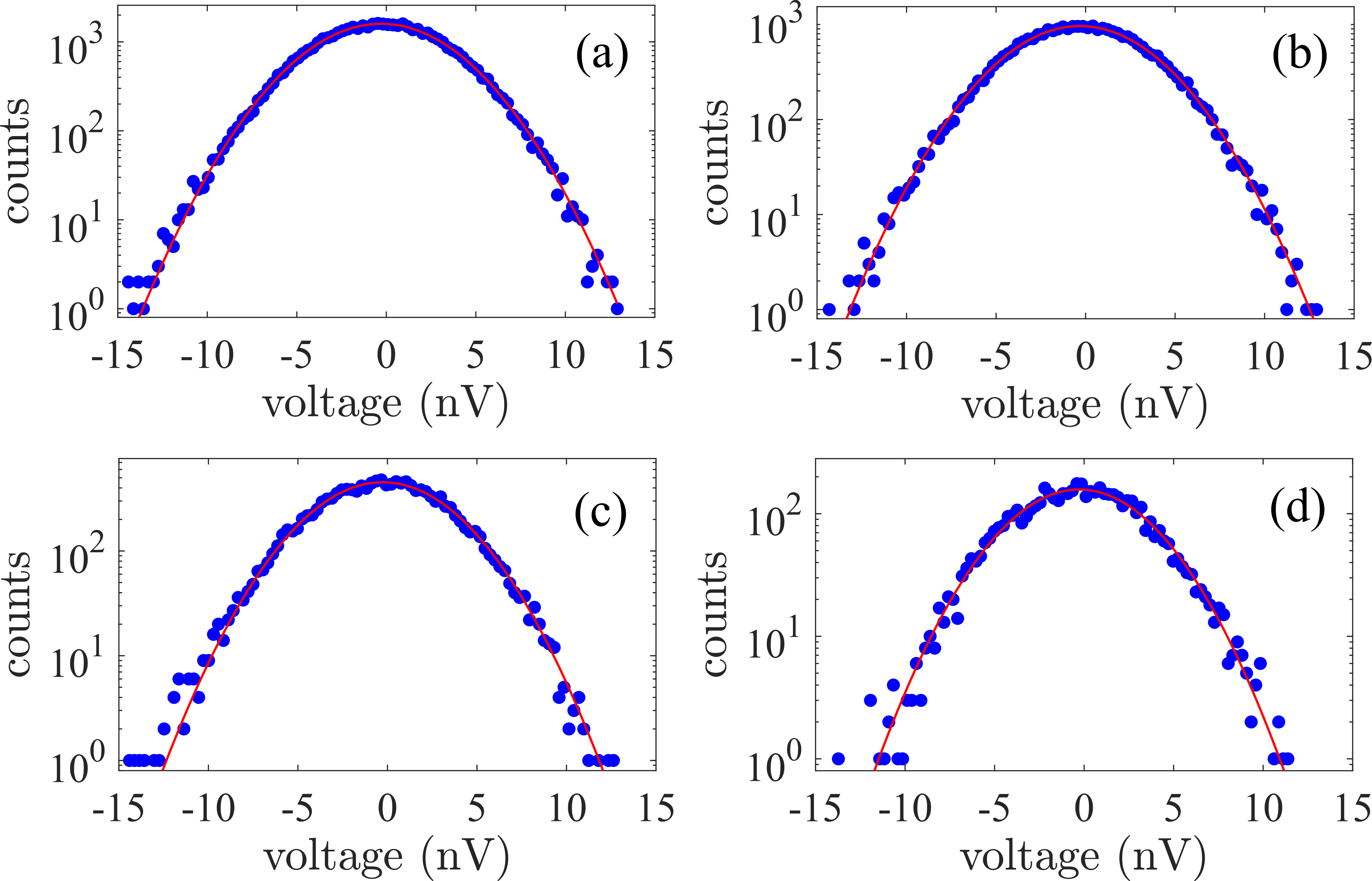}
	\caption{
		Histograms of the low voltage readings recorded during the experimental run. (a) Histogram of all the recorded low voltage readings (blinded analysis). (b) Histogram of the low voltage readings for control bits in the classically-generated bit string $c_1$. (c) Histogram of the low voltage readings for control bits in the bit string $q_2$, generated by IBM Quantum. (d) Histogram of the low voltage readings for control bits in the bit string $q_3$, generated by the NV center. All histograms are consistent with Gaussian distributions, shown as red lines.
	}
	\label{fig:2}
\end{figure}
The experimental run took place in November 2021. Our approach implemented a blinded measurement and data analysis procedure. The control computer software combined and randomly permuted the three bit strings, saving the permutation key for final unblinding. The resulting bit string controlled the output switch of the lock-in amplifier. The length of each switch cycle was $2\uu{s}$. In order to allow switching transients to decay, the computer recorded voltage data during the last second of each switch cycle. After linear drift subtraction and averaging, a single voltage reading was obtained for each switching cycle. Blinded data analysis divided these voltage readings into ``high'' and ``low'', based on whether they were greater or less than $1\uu{V}$. The histogram of the $49699$ low voltage readings is shown in fig.~\ref{fig:2}. The distribution is consistent with the Gaussian shape with mean $V_L = (-0.309\pm 0.015)\uu{nV}$. Here and below we quote the Gaussian standard error of the mean as the uncertainty. The average of the $51018$ high voltage readings is $V_H = (2.98\pm 8\times10^{-7})\uu{V}$. This matches the amplitude of the lock-in amplifier output waveform. The larger uncertainty for the mean $V_H$ value is due to the less-sensitive range used by the lock-in to make volt-level measurements, when the switch is closed. 

\begin{figure}[t]
	\centering
	\includegraphics[width=\columnwidth]{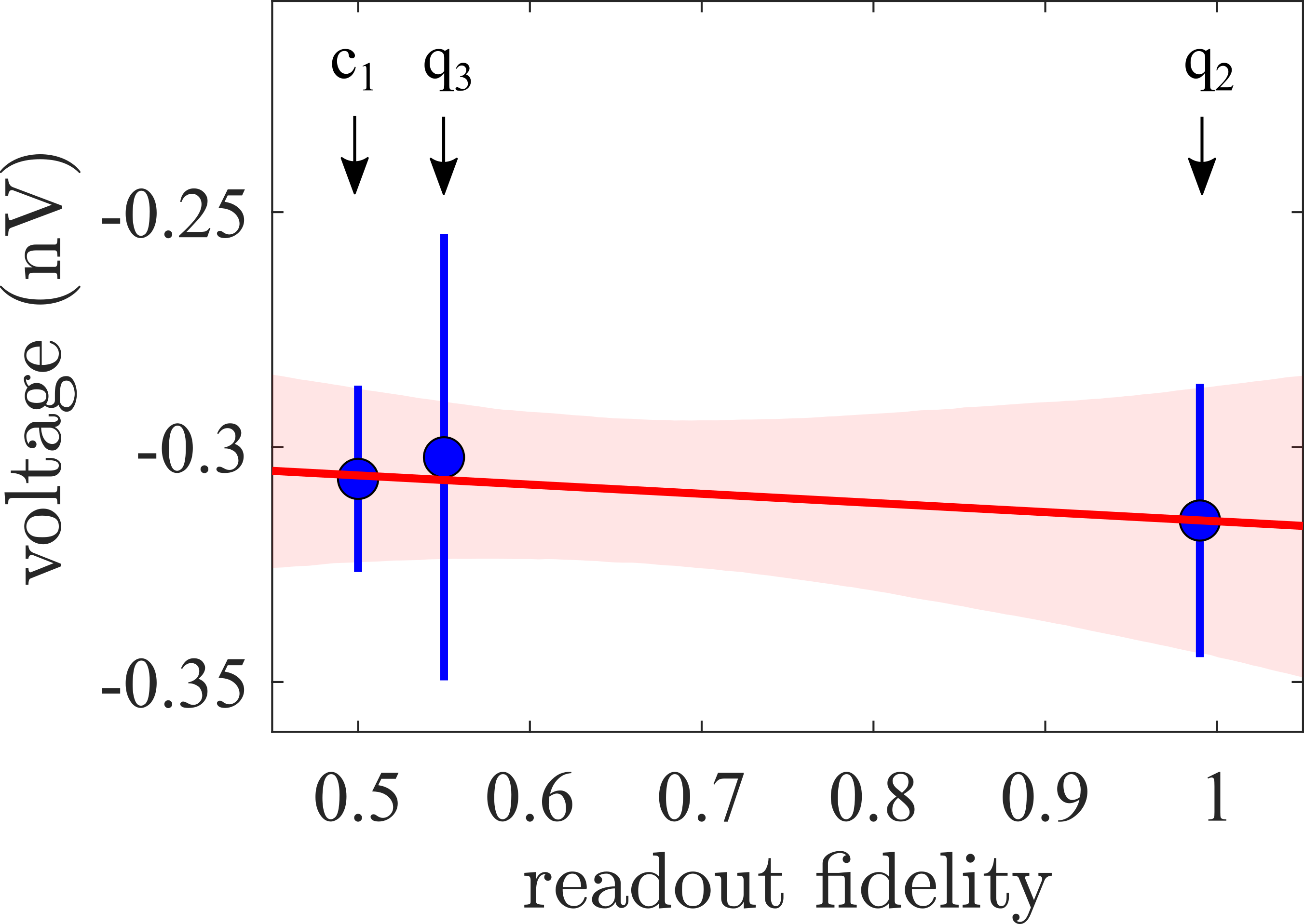}
	\caption{
		Detected open-switch voltage as a function of readout fidelity. Mean voltages obtained for each of the bit strings $c_1$, $q_2$, $q_3$ are marked with arrows. Each error bar corresponds to Gaussian standard error of the mean. The red line shows the best linear fit, and the shaded region indicates the one-standard-deviation uncertainties in the linear fit parameters, obtained using a Monte Carlo simulation with $10^4$ realizations.
	}
	\label{fig:3}
\end{figure}

After the analysis procedure was finalized, the bit permutation key was used to unblind the results. The voltage measurements for each of the three control bit strings were collected and analyzed separately. The results are shown in fig.~\ref{fig:3}. All three histograms are consistent with Gaussian-distributed data. The mean of the low voltage readings for bits from classically-generated bit string $c_1$ is $V_L^{(1)} = (-0.307\pm 0.020)\uu{nV}$. This is a measurement of the open-switch leakage voltage $V_s$. The mean of the low voltage readings for bits in string $q_2$, generated by the IBM Quantum processor, is $V_L^{(2)} = (-0.316\pm 0.029)\uu{nV}$. The mean of the low voltage readings for bits in string $q_3$, generated by the NV center in diamond, is $V_L^{(3)} = (-0.302\pm 0.047)\uu{nV}$. These values are plotted in fig.~\ref{fig:3}, with the corresponding readout fidelity on the horizontal axis. Linear regression gives the best-fit value of the open-switch leakage voltage: $V_s = (-0.306\pm0.019)\uu{nV}$. The quantum non-linearity parameter is extracted from the slope of the linear fit: $\epsilon_{\gamma} = (0.7\pm2.3)\times 10^{-11}$. Our measurements find no evidence for electromagnetic quantum non-linearity, and we set a 90\% confidence bound at the level $|\epsilon_{\gamma}|<4.7\times 10^{-11}$.

As discussed in ref.~\cite{Kaplan2021_}, the observable effects of non-linear quantum mechanics may be extremely sensitive to the full unknown cosmic history of the quantum state $|\psi\rangle$. Our work makes the plausible assumption that the universe has dominantly evolved classically, with negligible quantum spread~\cite{Kaplan2021_}. Within the Everett many-worlds interpretation of quantum theory, our measurements place limits on the electromagnetic interaction between different branches of the universe, created by initializing the qubit into a superposition state~\cite{Everett1957}.
We note that the open-switch voltage $V_s$ could have a contribution due to some unknown physical effects that modify the system hamiltonian; however our approach is not sensitive to such effects, since we search for the difference between measurement results for the qubit in the superposition state and for a mixture of classical bits.

It may be possible to improve experimental sensitivity to the electromagnetic quantum non-linearity. One possible avenue is to extend our approach to a larger dynamic range of macroscopic electromagnetic fields, controlled by a qubit measurement outcome. For example, switching a Tesla-level magnetic field with detection by a SQUID sensor may lead to a significant sensitivity gain. Another possibility is to make use of an ion interferometer, where non-linear effects cause the Coulomb field of one arm of the interferometer to affect the phase on the other arm.
The possibility of a solution to the black hole information problem is a strong motivation for experimental searches for gravitational non-linearity, which could make use of an accelerometer, such as an atomic interferometer, to measure the gravity gradient created by a test mass, whose position is modulated based on a qubit measurement.
Given the importance of testing the foundations of quantum mechanics, as well as potential applications, such as solutions of NP-complete problems in polynomial time~\cite{Kaplan2021_,Abrams1998,Aaronson2005,Bennett2009,Bao2016}, there is a strong case to explore all available experimental avenues.

\begin{acknowledgments}

The authors acknowledge discussions with A.~Polkovnikov and O.~P.~Sushkov.
We acknowledge the use of IBM Quantum services for this work. The views expressed are those of the authors, and do not reflect the official policy or position of IBM or the IBM Quantum team. 
A.V.G. and A.O.S. acknowledge support from the John Templeton Foundation Grant No. 60049570 and the Simons Foundation Grant No. 641332.
S.R. and D.K. are supported in part by the U.S.~National Science Foundation (NSF) under Grant No.~PHY-1818899.   
This work was supported by the U.S.~Department of Energy (DOE), Office of Science, National Quantum Information Science Research Centers, Superconducting Quantum Materials and Systems Center (SQMS) under contract No.~DE-AC02-07CH11359. 
S.R.~is also supported by the DOE under a QuantISED grant for MAGIS, and the Simons Investigator Award No.~827042.

\end{acknowledgments}


%

\end{document}